\documentclass[12pt, centerh1]{article}
\usepackage{colonequals,color,multirow,natbib}
\usepackage{graphicx}

\usepackage{amsfonts, amsmath, marvosym,amssymb}
\usepackage{algorithm}
\usepackage{bm,relsize}
\usepackage{geometry}

\usepackage[utf8]{inputenc}
\usepackage{lscape}
\usepackage{blkarray}
\usepackage{mathtools}
\usepackage{caption}
\usepackage{hyperref}
\usepackage{enumerate}
\usepackage{float}
\usepackage[section]{placeins}
\usepackage{subcaption}
\usepackage[T1]{fontenc}
\usepackage{tabularx}
\def\bzero{\boldsymbol{0}}

\def\bx{\mathbf{x}}
\def\tx{x}
\def\by{\mathbf{y}}
\def\ty{y}
\def\bw{\mathbf{w}}

\def\bz{\mathbf{z}}
\def\tz{z}

\def\bY{\mathbf{Y}}
\def\bI{\mathbf{I}}

\def\bc{\mathbf{c}}
\def\tc{c}

\newcommand{\bs}{\boldsymbol}

\newcommand{\E}{\mathbb{E}}
\DeclareMathOperator{\Tr}{tr}
\newcommand{\mvn}{\text{MVN}}

\begin{document}
\title{Robust Model-Based Clustering of Voting Records}

\author{Yang Tang\thanks{Department of Mathematics \& Statistics, McMaster University, Hamilton, ON, Canada.}\qquad\qquad Paul D.\ McNicholas$^*$ \qquad\qquad Antonio Punzo \thanks{Dept.\ of Economics and Business, University of Catania, Italy.}}
\date{}

\maketitle
\begin{abstract}
\noindent We explore the possibility of discovering extreme voting patterns in the U.S. Congressional voting records by drawing ideas from the mixture of contaminated normal distributions. A mixture of latent trait models via contaminated normal distributions is proposed. 
We assume that the low dimensional continuous latent variable comes from a contaminated normal distribution and, therefore, picks up extreme patterns in the observed binary data while clustering. We consider in particular such model for the analysis of voting records. The model is applied to a U.S. Congressional Voting data set on 16 issues. 
Note this approach is the first instance within the literature of a mixture model handling binary data with possible extreme patterns.  \\[-6pt]

\noindent \textbf{Keywords}:
Clustering, contaminated normal distribution; extreme patterns detection; latent variables; mixture models; U.S. Congressional Voting.
\end{abstract}

\section{Introduction}
\label{sec:Intro}
The United States has amassed a vast global influence since the late 19th century. Dozens of developing countries look to external examples for best practices when handling complex matters, and the U.S. is the obvious first port of call. The U.S. has the oldest working national constitution in the world, as well as strong institutions and rule of law to accompany it. The United States Constitution grants all legislative power to the Congress, which consists of a Senate and a House of Representatives. Prior to the president signing a bill, it must be passed by the Senate and the House of Representatives. 
Therefore, legislative voting records are often used to make inferences about the preferences of legislators, issues that drive political conflict, the cohesiveness of parties, and the existence of cooperation across parties. For example, a spending bill rejected by senators on January 19, 2018 is evidence that partisanship in the United States Congress is at an historic high. 
\begin{quote}
``The federal government shut down for the first time in more than four years Friday after senators rejected a temporary spending patch and bipartisan efforts to find an alternative fell short as a midnight deadline came and went.''
-- The Washington Post
\end{quote}
Although party-line voting has become the new normal, there still exist individuals in the Senate and the House of Representatives who vote with members of the opposite party. For instance, thirteen Republicans voted against a major tax cut bill on November 16, 2017. 
\begin{quote}
``House Republicans passed major tax cut legislation on Thursday after a closed-door speech on Capitol Hill from Donald Trump.
The bill, which would slash corporate taxes from 35\% to 20\% and also reduce individual rates, passed by a margin of 227 to 205, with support from all but 13 Republicans and no Democrats.''
--The Guardian
\end{quote}

Statistical analyses of legislation and legislators provide context for the legislative process. New ways of analyzing voting behaviour in the U.S. Congress is a topic that has been receiving increasing attention over the past few years. The most popular legislator partisanship indexing methods \citep[e.g.,][]{poole84,poole97,cox02} put each member of Congress on a single (liberal-to-conservative) linear scale (i.e., dimension). These dimensions are considered valuable and are accepted as standard practice for quantifying polarization. However, the actual vote cocktail used to create the index, as well as how this value is transformed to a linear value, are very difficult to explain. Furthermore, indexing methods are described in whole by aggregate measures, such as mean of members' indexes, which can be affected by members who vote across party lines \citep[i.e., extreme voting patterns;][]{ornstein13}. \cite{andris15} propose a network model which leverages raw and disaggregate roll call votes from the U.S. House of Representatives to find the mutual agreement rates on legislative decisions among pairs of representatives. A threshold value of average pairwise agreements is defined to identify cooperative votes across party lines. This approach depends solely on the quantity of agreements; hence identifying the issues that drive political conflict is impossible.  

We are particularly interested in analyzing voting records from the House of Representatives because each representative is from one of the 435 unique Congressional districts. The voting patterns from each of them should reflect the distinctive perspective of his or her unique district and constituency. Our goal herein is to separate voting patterns into meaningful clusters as well as detect representatives who vote across party lines \citep[which has not been a very common occurrence in the last few decades;][]{andris15}.
  
Mixture model-based clustering is recently receiving broad interest. A mixture model is a convex linear combination of a finite number of component distributions, and according to this approach, data are clustered using some assumed mixture modelling structure. The density of a general finite mixture model, for the generic $i$th observation $ \bx_i$, is given by
\begin{equation}\label{eq:fmm}
f(\bx_i;\boldsymbol \Theta) = \sum_{g=1}^G \pi_gp\left(\bx_i; \boldsymbol{\theta}_g\right),
\end{equation}
where $G$ is the number of components, $\pi_g\in(0,1]$, such that $\sum_{g=1}^G\pi_g=1$, is the $g$th mixing proportion, 
and $\boldsymbol \theta_g$ contains unknown parameters in the mixture model.
For non-continuous data, one needs to specify $p(\bx_i; \boldsymbol \theta_g)$ in \eqref{eq:fmm} through probability mass functions. While there are plenty of choices for univariate non-continuous distributions, the use of multivariate non-continuous distributions for the definition of the mixture models is limited due to the difficulty in constructing models that allow flexibility of the dependence structure. Mixture models with latent structure have been considered for the analysis of non-continuous data and mixed type data. The latent trait models use a continuous univariate or multivariate latent variable to model the dependence among the categorical response variables (e.g., raw voting records). Recently proposed, mixtures of latent trait models for the analysis of categorical and mixed response variables include work by \citet{muthen06}, \citet{vermunt07}, \citet{khan10}, \citet{browne12}, \citet{cagnone12}, \citet{gollini14}, and \cite{tang15}. \citet{gollini14} propose a mixture of latent trait analyzers (MLTA) for model-based clustering of binary data, wherein a categorical latent variable identifies clusters of observations and a latent trait is used to accommodate within-cluster dependency. When applied to voting records from the U.S. House of Representatives on sixteen key issues in 1984, the MLTA finds four groups and the Democrat representatives are divided into two voting blocs. This is because voting patterns that are least likely to occur under the hypothesized model (referred to as extreme patterns herein) can affect the estimation of model parameters and therefore the group membership. However, model-based approaches for handling binary data with possible extreme patterns have received relatively little attention. To achieve our goal of automatically detecting extreme patterns, we propose a mixture of latent trait model via contaminated normal distributions. 
This is the first instance of a mixture model handling binary data with possible extreme patterns.

The rest of this article is organized as follows. 
The model-based clustering framework is outlined in Section~\ref{sec:framework}, and an expectation-conditional maximization (ECM) algorithm for parameter estimation is described. 
Model selection and performance assessment are detailed in Section~\ref{sec:modelselection}. 
Application on artificial data and the U.S. Congressional Voting data are presented in Section~\ref{sec:application} and we conclude with some discussion in Section~\ref{sec:conclusion}.

\section{Methodology}
\label{sec:framework}

Let $\bx_i=\left(\tx_{i1},\ldots,\tx_{iM}\right)$ be a sequence of $M$ binary observed response variables for the $i$th statistical unit. 
We assume that each observation $\bx_i$, $i = 1,\ldots,n$, comes from one of the $G$ components of a mixture model and we use $\bz_i=\left(\tz_{i1},\ldots,\tz_{iG}\right)$ to identify the group memberships, where $\tz_{ig}=1$ if observation $\bx_i$ is in component $g$ and $\tz_{ig}=0$ otherwise.

\subsection{Mixture of Latent Trait Models} 
\label{sec:MLTA}

\citet{gollini14} assume that the conditional distribution of $\bx_i$ given that the observation is from group $g$ (i.e., $\tz_{ig}=1$) is a latent trait model with the intercept parameters $\alpha_{mg}$ and the slope parameters $\bw_{mg}$.
Their mixture of latent trait analyzers (MLTA) model assumes there is a $D$-dimensional continuous latent variable $\bY$ underlying the behavior of the $M$ binary response variables, where $\bY_i\sim \mvn\left(\bzero,\bI\right)$. 
Thus, the MLTA model is of the form
\begin{equation*}
p(\bx_i)=\sum_{g=1}^G\pi_g \int\limits_{\by_i} p(\bx_i|\by_i, \tz_{ig}=1)p(\by_i)d\by_i,
\end{equation*}
where 
\begin{equation*} 
p(\bx_i|\by_i,\tz_{ig}=1)=\prod_{m=1}^M \{p_{mg}(\by_i)\}^{\tx_{im}}\{1-p_{mg}(\by_i)\}^{(1-\tx_{im})},
\end{equation*}
and the response function for each group is given by
\begin{equation*}  
p_{mg}(\by_i)=p(\tx_{im}=1|\by_i,\tz_{ig}=1)=\frac{1}{1+\exp\{-(\alpha_{mg}+\bw_{mg}'\by_i)\}}.
\end{equation*}

In particular, $\alpha_{mg}$ has a direct effect on the probability of a positive response to the variable $m$ given by an individual in group $g$, 
\begin{equation*}  
p_{mg}(\bzero)=p(\tx_{im}=1|\by_i=\bzero, \tz_{ig}=1)=\frac{1}{1+\exp(-\alpha_{mg})}.
\end{equation*}
The value $p_{mg}(\bzero)$ is the probability that the median individual in group $g$ has a positive response for the $m$th variable. 
The value of the slope parameters $\bw_{mg}$ account for the correlation between the observed response variables.
The log-likelihood can be written as,
\begin{equation}  \label{ch2:eq:mltadensity}
l=\sum_{i=1}^n\log\left\{\sum_{g=1}^G \pi_g \int \limits_{\by_i} \prod_{m=1}^M \, p(\tx_{im}|\by_i, \tz_{ig}=1)p(\by_i)d\by_i\right\}.
\end{equation}

\subsection{Contaminated Normal Distribution}
\label{sec:CG}

A contaminated normal distribution is a two-component normal mixture in which one of the components -- with a large prior probability -- represents normal observations, and the other -- with a small prior probability, the same mean and a proportionally inflated covariance matrix -- represents the extreme points \citep{aitkin80}. For continuous multivariate random variables, the mixture of contaminated normal distributions accommodates outlying observations, spurious observations or noise, and a family of fourteen parsimonious mixtures of contaminated normal distributions has been developed for robust clustering \citep{punzo16a}. 
     The proposed mixture of latent trait models via a contaminated normal distribution can be used to model multivariate binary (categorical) observations.


\subsection {Mixture of Latent Trait Model via Contaminated Normal Distributions}\label{sec:MLTCG}
The latent trait models via a contaminated normal distribution assume that there is a $D$-dimensional continuous latent variable $\bY_i=\left(\ty_{i1}, \ldots, \ty_{iD}\right)$ underlying the behaviour of the $M$ binary response variables within each observation $\bx_i$  
\begin{equation*}
p(\bx_i|\bs \Theta)=\int \limits_{\by_i}p(\bx_i|\by_i; \bs \alpha, \bw)p(\by_i| \tau, \eta)d \by_i,
\end{equation*}
where the conditional distribution of $\bx_i$ given $\by_i$ is
\begin{equation*}
p(\bx_i|\by_i)=\prod_{m=1}^M \{p_m(\by_i)\}^{\tx_{im}}\{1-p_m(\by_i)\}^{(1-\tx_{im})},
\end{equation*}
and the response function is a logistic function 
\begin{equation*}
p_m(\by_i)=p(\tx_{im}=1|\by_i)=\frac{1}{1+\exp\{-(\alpha_m+\bw_m'\by_i)\}},
\end{equation*}
where $\alpha_m$ is the intercept and $\bw_m$ are the slope parameters in the logistic function.
The continuous latent variable  $\bY_i$ comes from a contaminated normal distribution 
\begin{equation*}
p(\by_i\mid\tau, \eta)=\tau p(\by_i \mid \tc_i=1)+(1-\tau)p(\by_i \mid \tc_i=0),
\end{equation*}
Such distribution can be hierarchically represented as
\begin{align*}
& \bY_i\mid \tc_i=1 \sim \mvn(\bs 0, \bI), \\
& \bY_i\mid \tc_i=0  \sim \mvn(\bs 0, \eta \bI),
\end{align*}
where $\tau\in(0.5,1)$ is the prior probability of a randomly chosen $\bY_i$ coming from $ \mvn(\bs 0, \bI )$ and $\eta>1$ denotes the degree of contamination. Because of the assumption $\eta>1$, it can be interpreted as the increase in variability due to the extreme values. 

The mixture of latent trait model via contaminated normal distributions (MLTCN) is a mixture latent trait model and the latent variables are random variables from a contaminated normal distribution with density
\begin{equation}\label{eq:mltcg}
p(\bx_i)=\sum_{g=1}^G\pi_g\int \limits_{\by_{ig}} p(\bx_i|\by_{ig};\bs \alpha_g, \bw_g)p(\by_{ig}|\tau_g,\eta_g) d\by_{ig}, 
\end{equation}
where $\pi_g$, $\bs \alpha_g$, $\bw_g$, $\tau_g$, and $\eta_g$ are component specific parameters.


\subsection{Model Fitting}
\label{sec:fitting}
In the mixture of latent trait models, the integral in \eqref{ch2:eq:mltadensity} and \eqref{eq:mltcg} is intractable. 
A variational approach can be used for a fast algorithm since it has a closed-form solution for parameter updates and provides a lower bound approximation to the log-likelihood. The variational parameters $\bs \xi_{ig}=(\xi_{i1g},\ldots,\xi_{iMg})$ are introduced to approximate the logarithm of the component densities with a lower bound \citep[see][]{gollini14}. 
To fit the MLTCN model, we adopt the variational expectation-conditional maximization (ECM) algorithm. 
The ECM algorithm \citep{meng93} replaces a complex M-step by a number of computationally simpler conditional maximization steps (CM-steps) and the variational approximation is used to obtain a lower bound for the log-likelihood. 

In our model, there are two sources of incomplete data: one arises from the fact that we do not know the component labels $\bz_i$ and the other arises from the fact that we do not know whether an observation in group $g$ is normal or extreme. 
To denote the second source of missing data, we use $\bc_i=\left(\tc_{i1}, \ldots , \tc_{iG}\right)$ where $\tc_{ig}=1$ if observation $i$ in group $g$ is normal and $\tc_{ig}=0$ if observation $i$ in group $g$ is extreme. 
Therefore, the complete-data log-likelihood can be written as 
\begin{equation*}
\begin{split}
&l_c=\sum_{i=1}^{n}\sum_{g=1}^G \tz_{ig}\left\{\log\pi_g +\tc_{ig}\log p(\bx_i|\by_{ig1}; \bs \alpha_g, \bw_g)+\tc_{ig}\log\tau_g+c_{ig}\log p(\by_{ig1})\right.\\
&\left.+(1-\tc_{ig})\log(1-\tau_g)+(1-\tc_{ig})\log p(\bx_i|\by_{ig0}; \bs \alpha_g, \bw_g)+(1-c_{ig})\log p(\by_{ig0})\right\},
\end{split}
\end{equation*}
where $\bY_{ig1}\sim\mvn(\bs 0, \bI)$ and $\bY_{ig0}\sim\mvn(\bs 0, \eta_g\bI)$.

The ECM algorithm iterates between three steps, an E-step and two CM-steps, until convergence. The parameter vector $\bs \theta_g$ is partitioned in $\bs \theta_g=\{\bs \theta_{g1}, \bs \theta_{g2}\}$, where $\bs  \theta_{g1}=\{\boldsymbol \xi_g\}$ and $\bs \theta_{g2}=\{\bs \alpha_g, \bw_g, \eta_g, \pi_g, \tau_g \}$. The steps of the ECM algorithm, for the generic $(t + 1)$th iteration, $t = 1, 2, \ldots$, are detailed in \appendixname~\ref{EMdetail}.

\subsection{Convergence Criterion}
\citet{bohning94} exploited Aitken's acceleration procedure \citep{aitken26} as a convergence criterion. This stopping criterion determines convergence by estimating the limiting value of the log-likelihood at each iteration of the ECM algorithm. The Aitken acceleration at iteration $t$ is
$$a^{(t)}=\frac{l^{(t+1)}-l^{(t)}}{l^{(t)}-l^{(t-1)}},$$
where $l^{(t)}$ is the log-likelihood at iteration $t$. An asymptotic estimate of the log-likelihood at iteration $t$ is
$$l_{\infty}^{(t)}=l^{(t-1)}+\frac{1}{1-a^{(t-1)}}(l^{(t)}-l^{(t-1)}).$$
\citet{bohning94} suggest considering an algorithm to be converged when
$|l_{\infty}^{(t+1)}-l_{\infty}^{(t)}|<\epsilon,$ and we use this criterion herein with $\epsilon=0.01$.

\section{Model Selection and Performance Assessment}
\label{sec:modelselection}

\subsection {Model Selection}\label{sec:BIC}

The Bayesian information criterion \citep[BIC;][]{schwarz78} is commonly used for model selection in model-based clustering. The BIC takes the form, 
\begin{equation}\label{eqn:bic}
\text{BIC}=-2l+k\log n,
\end{equation}
where $l$ is the maximized log-likelihood, $k$ is the number of free parameters to be estimated in the model, and $n$ is the number of observations. \citet{schwarz78} proved that, if one of the models is correct, so that there is a true $\bs \Theta$ in that model, as $n$ becomes large, with probability approaching 1, BIC will select the best model. \citet{poskitt87} and \citet{haughton88} extended and improved Schwarz's work, showing that consistency held also under less restrictive conditions. 
The BIC is used as a model selection criterion to select the number of clusters $G$ and the dimension $D$ of the latent variable $\bY$. 
When defined as in \eqref{eqn:bic}, models with lower values of BIC are preferable. 

\subsection{Performance Assessment}

When the true underlying groups are known, as in the simulations (Section~\ref{sec:sumulation}), the clustering performance of the models can be assessed by comparing the known class labels to the estimated group memberships. We assign each observation $\bx_i\in \{\bx_1, \ldots, \bx_n\}$ to one and only one group $g$, $g=1, \ldots, G$, with the largest corresponding $\tz_{ig}$ value after convergence. These are referred to as the MAP classifications. 
The Rand index \citep{rand71} is simply calculated as pairwise agreements between the true class labels and the MAP classification, 
\begin{equation}
\label{eqn:rand}
\frac{\text{number of pairwise agreements}}{\text{total number of pairs}}.
\end{equation}
The Rand index lies between $0$ and $1$.
When two partitions agree perfectly, the Rand index is $1$. A problem with the Rand index is that the expected value of the Rand index of two random partitions does not take a constant value (say zero) and smaller values are difficult to interpret. 
Therefore, \citet{hubert85} proposed the adjusted Rand index (ARI). 
The ARI is the corrected-for-chance version of the Rand index. The general form of the ARI is 
$$
\frac{\text{index}-\text{expected index}}{\text{maximum index}-\text{expected index}},
$$ 
which is bounded above by 1, and has expected value 0 under random classification.

\section{Data Analysis}
\label{sec:application}

In this section, we evaluate the performance of the MLTCN model on artificial data and the U.S. Congressional Voting data.  Particular attention will be devoted to the problem of contamination parameter recovery and model selection in the simulation study (Section~\ref{sec:sumulation}), clustering results, parameter interpretation and detecting extreme patterns in the application of real data (Section~\ref{sec:Voting}).

\subsection{Simulation Studies} 
\label{sec:sumulation}

To illustrate the ability of parameter recovery for the proposed MLTCN model, we perform a simulation experiment on a $25$-dimensional binary data set (i.e.,~$M=25$). 
The observations are generated from a MLTCN model with a two-component mixture ($G=2, \pi_1=\pi_2=0.5$). 
The latent variable is from a two-dimensional (i.e.,~$D=2$) multivariate normal distribution. 
Each component consists of $80\%$ normal patterns and $20\%$ extreme patterns (i.e., $\tau_1=\tau_2=0.8$) and degrees of contamination $\eta_1=\eta_2=2.5$. 
We choose sample sizes $n \in \{100,200,500\}$ and run 100 simulations for each sample. Data were fitted using $G=2$ and $D=2$, and starting randomly. 
\tablename~\ref{ch5:table:3.1} presents the value of the estimated contamination parameters and standard errors of these estimates for $n \in \{100,200,500\}$.  
The standard errors are relatively low when $n=100$ and decrease with increasing sample size $n$.

\begin{table}
\caption{\label{ch5:table:3.1}Estimated values for $\bs\eta$ and $\bs \tau$.} 
	\centering
\scalebox{1}{
\begin{tabular*}{1\textwidth}{@{\extracolsep{\fill}}rcccccccc}	
\hline
&\multicolumn{2}{c}{$n=100$}&&\multicolumn{2}{c}{$n=250$}&&\multicolumn{2}{c}{$n=500$}\\
\cline{2-3}\cline{5-6}\cline{8-9}
&$\eta_g $ &$\tau_g$ &&$\eta_g $ &$\tau_g$ &&$\eta_g $ &$\tau_g$\\
&(SE)&(SE)&&(SE)&(SE)&&(SE)&(SE)\\ [0.5ex] 
\cline{2-9}
 \hspace{0.2cm}\multirow{2}{*}{Group 1}&$2.37$&$0.79$&&$2.51$&$0.80$&&$2.51$&$0.80$\\
&(0.25)&(0.03)&&(0.10)&(0.01)&&(0.08)&(0.01)\\
\multirow{2}{*}{Group 2} &$2.45$&$0.80 $&&$2.50$&$0.80$&&$2.50$&$0.80$\\
&(0.22)&(0.03)&&(0.10)&(0.01)&&(0.07)&(0.01)\\
\hline
\end{tabular*}}
\end{table}

\begin{figure}
	\centering
	\makebox{\includegraphics[width=0.6\textwidth]{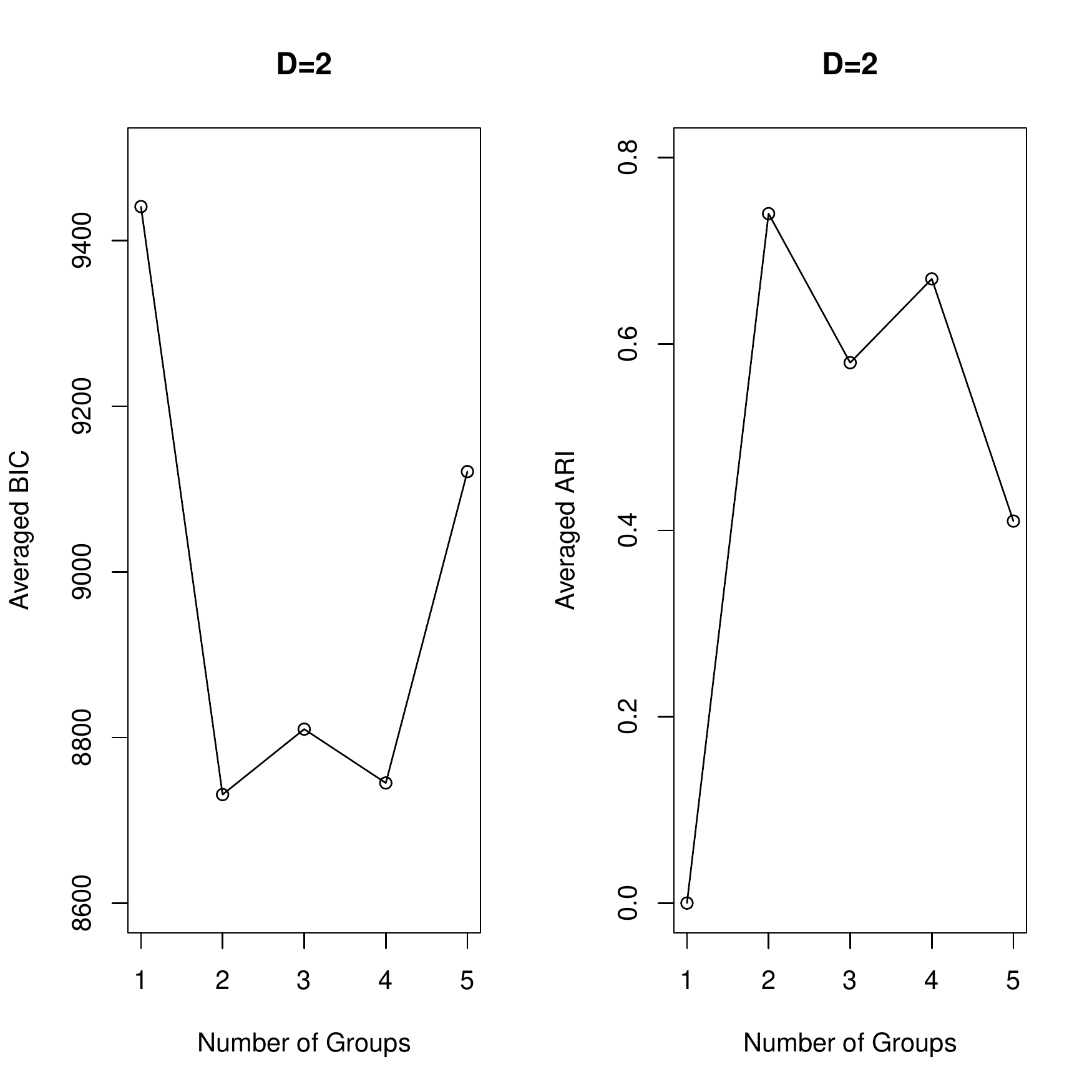}}
	\caption{\label{ch5:fig:3.1}BIC values averaged on the 100 samples for each value of $G$ (left), and ARI averaged on the 100 samples for each value of $G$ (right).}
\end{figure}

The samples with $500$ observations were fitted using $G \in \{1,2,3,4,5\}$ and $D=2$. The left panel of \figurename~\ref{ch5:fig:3.1} displays the BIC values averaged on the 100 samples for each value of $G$. As shown in the right panel of \figurename~\ref{ch5:fig:3.1}, on average the ARI has a maximum for $G=2$. This result is  an evidence of the BIC selecting the ``best'' model.  

\subsection{U.S. Congressional Voting} 
\label{sec:Voting}
The U.S. Congressional Voting data includes votes of 435 members ($n=435$) of the U.S.\ House of Representatives on sixteen key issues ($M=16$) in 1984 with three different types of votes: yes, no, or undecided. The representative's party is labeled as a Democrat or a Republican. The issues voted on are listed in \tablename~\ref{table:add}.
There are $11\%$ undecided votes for issue 2 and $23\%$ for issue 16. All other issues have less than $5\%$ undecided votes. 
We code each question in two binary variables A and B: the responses for the A
variables are coded as 1 = yes/no and 0 = undecided; and B variables are 1 = yes, 0 =no/undecided.
\begin{table}
	\caption{\label{table:add}The issues voted on in the U.S.\ Congressional Voting data.}
	\centering
	\fbox{%
		\begin{tabular*}{1\textwidth}{@{\extracolsep{\fill}}cp{5cm}cp{4cm}}
			Item &Issue & Item  & Issue\\
			\hline
			1  &Handicapped Infants &  9 &MX Missile\\
			2  & Water Project Cost-Sharing & 10 & Immigration \\
			3  & Adoption of the Budget Resolution &11& Synfuels Corporation Cutback\\
			4& Physician Fee Freeze& 12 & Education Spending\\
			5  & El Salvador Aid  &13 & Superfund Right to Sue\\
			6 & Religious Groups in Schools  &14 &Crime\\
			7  &Anti-Satellite Test Ban &15 & Duty- Free Exports\\
			8 & Aid to Nicaraguan `Contras' & 16 & Export Administration Act/South Africa\\
	\end{tabular*}}
\end{table}

We compare our results to those obtained by fitting a MLTA model and a MCLT model \citep{tang15}. 
The MLTCN was fitted to these data for  $D=1,2,3,4$ and $G=1,2,3,4$. 
The minimum BIC (\tablename~\ref{ch5:table:3.2}) occurs at the 2-component, 2-dimensional model. 
The BIC value is $9918$. 
\begin{table}
\caption{\label{ch5:table:3.2}The estimated BIC for the models with $D=1,2,3,4$ and $G = 1,2,3,4$.}
\centering
\scalebox{1}{
\begin{tabular*}{1\textwidth}{@{\extracolsep{\fill}}r c c c c }
\hline
&$G=1$&$G=2$&$G=3$&$G=4$\\ 
\hline
\hspace{0.2cm}$D=1$&10509&11480&10282&10558\\
$D=2$&10557&\textbf{9918}&10305&10713\\
$D=3$&12176 &10282&10712&11364\\ 
$D=4$&12258&10612&11354&12051\\
\hline
\end{tabular*}}
\end{table}

A summary of the best models for the MLTA, PMLTA, MCLT and MLTCN approaches is shown in \tablename~\ref{ch5:table:3.3}. It can be seen that the highest ARI value ($0.77$) is obtained using the MLTCN model.  
\begin{table}
\caption{\label{ch5:table:3.3}A comparison of 4 different approaches.}
\centering
\scalebox{1}{
\begin{tabular*}{1\textwidth}{@{\extracolsep{\fill}}r l c c c c c}
\hline
&Model&$G$&$D$&BIC&$\boldsymbol{\Sigma}_g$&ARI\\ 
\hline
\hspace{0.2cm}1&MLTA&$3$&$1$&9812&n/a&$0.42$\\
2&PMLTA&$4$&$2$&$9681$&n/a&$0.47$\\
3&MCLT&$2$&$5$&9597&EVI&0.64 \\ 
4&MLTCN&$2$&$2$&$9918$&n/a& \textbf{0.77}\\
\hline
\end{tabular*}}
\end{table}%

The classification table for group membership versus party membership for the selected model ($G=2$, $D=2$) is presented in \tablename~\ref{ch5:table:3.4}. In comparison with the true party membership, there are only 26 misclassified representatives (i.e., 94.02\% accuracy) associated with the chosen model. Group~1 consists mainly of Republican representatives, and Group~2 consists mainly of Democratic representatives. Due to the number of variables, it is difficult to know the possible presence of extreme patterns. The selected MLTCN model recognizes the presence of the two groups when we consider the normal points together with the extreme points. The advantage of our approach is that not only can we cluster in the presence of extreme patterns, but we can also identify them. When we view the results of our analysis, we see that there are 161 extreme observations, and it is not surprising that 20 out of 26 misclassified observations are considered extreme observations. 

\begin{table}
\caption{\label{ch5:table:3.4}Cross-tabulation of the parties and predicted classification for our chosen model ($G=2$, $D=2$) for the U.S.\ Congressional Voting data.}
\centering
\scalebox{1}{
\begin{tabular*}{1\textwidth}{@{\extracolsep{\fill}}c rrrr  }
\hline
&Group 1&Normal/Extreme&Group 2&Normal/Extreme \\ 
\hline
\multirow{ 2}{*}{Republican}& \multirow{ 2}{*}{7}&1 &\multirow{ 2}{*}{161}&118\\
&&6&&43\\
\hline
\multirow{ 2}{*}{Democrat}&\multirow{ 2}{*}{248}&150& \multirow{ 2}{*}{19}&5\\  
&&98&&14\\
\hline
\end{tabular*}}
\end{table}

\tablename~\ref{ch5:table:3.5} and \ref{ch5:table:3.6} shows the median probability $p_{mg}(\bzero)$ for each of the clusters. The probabilities of positive responses for the A variables (yes/no vs.\ undecided) for the median individuals in all clusters are always high with only one exception in the normal observations in Group~1, for variable number 16, where $p_{16\,1}(\bzero)=0.37$. Thus, the majority of representatives voted on most issues, but with a slightly higher voting rate in extreme observations on all issues. Due to the high voting rates, most probabilities given for B variables (yes vs.\ no/undecided) can be interpreted in terms of voting yes versus no.

It can be observed that the responses for the median individual in Group 1 are opposite to the ones given by the median individual in Group 2 for most issues. The extreme observations in Group 1 showed different voting behaviour on Issue 5 (El Salvador Aid), 9 (MX Missile) and 16 (Export Administration Act/South Africa) (\tablename~\ref{ch5:table:3.5}). The extreme observations in Group 2 showed different voting behaviour on Issue 7 (Anti-Satellite Test Ban), 12 (Education Spending), 13 (Superfund Right to Sue), and 16 (Export Administration Act/South Africa); see \tablename~\ref{ch5:table:3.6}.

\begin{table}
	\caption{\label{ch5:table:3.5}A comparison of the probability of a positive response for individuals classified as ``Normal'' vs. ``Extreme'' in Group 1.}    
	\centering
\scalebox{0.85}{
		\begin{tabular*}{1.07\textwidth}{@{\extracolsep{\fill}}cccccc}
		\hline
			Y/N vs.\ Undecided   & Normal    & Extreme     & Y vs.\ N/Undecided      & Normal    & Extreme \\
			\hline
			1A    &      0.97  &      0.98     & 1B    &      0.63  &      0.58  \\
			2A    &      0.89  &      0.92         & 2B    &      0.54  &      0.35  \\
			3A    &      0.98  &      0.98         & 3B    &      0.88  &      0.91  \\
			4A    &      0.96  &      0.99         & 4B    &      0.03  &      0.05  \\
			5A    &      0.96  &      0.99         & 5B    &       \textbf{0.33}  &    \textbf{0.00} \\
			6A    &      0.94  &      0.99         & 6B    &      0.56  &      0.30  \\
			7A    &      0.96  &      1.00         & 7B    &      0.64  &      0.93  \\
			8A    &      0.98  &      0.99         & 8B    &      0.76  &      0.98 \\
			9A    &      0.86  &      1.00         & 9B    &      \textbf{0.51}  &      \textbf{ 0.96}  \\
			10A   &      0.98  &      1.00         & 10B   &      0.39  &      0.59  \\
			11A   &      0.94  &      0.99         & 11B   &      0.54  &      0.39  \\
			12A   &      0.93  &      0.95         & 12B   &      0.18  &      0.05  \\
			13A   &      0.96  &      0.95         & 13B   &      0.37  &      0.14  \\
			14A   &      0.96  &      0.98         & 14B   &      0.36  &      0.27  \\
			15A   &      0.93  &      0.97         & 15B   &      0.57  &      0.69  \\
			16A   &       \textbf{0.37}  &       \textbf{1.00}       & 16B   &       \textbf{0.33}  &  \textbf{1.00} \\
			\hline
	\end{tabular*}}
\end{table}

\begin{table} 
	\caption{\label{ch5:table:3.6}A comparison of the probability of a positive response for individuals classified as ``Normal'' vs. ``Extreme'' in Group 2.}  
	\centering
 \scalebox{0.85}{
		\begin{tabular*}{1.07\textwidth}{@{\extracolsep{\fill}}cccccc}
			\hline
			Y/N vs.\ Undecided   & Normal    & Extreme    & Y vs.\ N/Undecided      & Normal    & Extreme \\
			\hline
			1A    &      0.95  &      1.00        & 1B    &      0.14  &      0.29  \\
			2A    &      0.85  &      0.92         & 2B    &      0.54  &      0.25  \\
			3A    &      0.95  &      1.00         & 3B    &      0.06  &      0.33  \\
			4A    &      0.96  &      1.00         & 4B    &      0.93  &      0.92  \\
			5A    &      0.96  &      1.00         & 5B    &      0.94  &      0.98 \\
			6A    &      0.98  &      1.00         & 6B    &      0.95  &      0.81  \\
			7A    &      0.93  &      1.00         & 7B    &      \textbf{0.04}  &  \textbf{0.67}  \\
			8A    &      0.92  &      0.98         & 8B    &      0.03  &      0.33 \\
			9A    &      0.96  &      1.00         & 9B    &      0.03  &      0.27  \\
			10A   &      0.97  &     1.00         & 10B   &      0.44  &      0.67 \\
			11A   &      0.90  &      0.98         & 11B   &      0.17  &      0.17  \\
			12A   &      0.89  &      0.96        & 12B   &     \textbf{0.17}  &   \textbf  {0.77} \\
			13A   &      0.92  &      0.94        & 13B   &     \textbf{0.89}  &     \textbf{0.58}  \\
			14A   &      0.93  &     1.00        & 14B   &      0.91  &      1.00 \\
			15A   &      0.91  &      0.94         & 15B   &      0.02  &      0.21  \\
			16A   &     0.80  &      1.00       & 16B   &     \textbf{0.36}  &     \textbf{1.00} \\
			\hline
	\end{tabular*}}
\end{table}

\section{Discussion}
\label{sec:conclusion}

The MLTCN model has been introduced for robust clustering of the U.S. Congressional Voting data. 
It can be viewed as a generalization of the MLTA that accommodates extreme patterns in binary data via contaminated normal distributions; specifically, it can automatically detect extreme observations (cooperators) while clustering. The MLTCN model is demonstrated to be effective in clustering.
 
Real data are often ``contaminated'' and it is difficult to detect extreme observations in high-dimensional binary data because the data cannot be easily visualized. 
When applied to the U.S. Congressional Voting data, our approach performed better in terms of classification when compared to the MLTA and MCLT models. The model parameters are interpretable and provide a characterization of the extreme observations. The result also showed that, in the $98$th Congress, Democrats and Republicans cooperate fairly often.    

Future work will focus on the development of {Python} code to improve computing time of our MLTCN model. 
In \citet{tang15} a parsimonious family of the mixture of latent trait models was developed by using common slope parameters and applying restrictions to the components of the decomposed covariance matrices. Analogous families of parsimonious models could be developed for the MLTCN model to further reduce the number of parameters to be estimated.

\appendix
\section{Detailed Parameter Estimation}\label{EMdetail}
The steps of the ECM algorithm, for the $(t + 1)$th iteration, $t = 1, 2, \ldots$, are detailed below.

\hspace{-0.5cm}(a) Estimate $z_{ig}$ and $c_{ig}$
\begin{equation*}
\begin{split}
\tz_{ig}^{(t+1)}=&\frac{\pi_g^{(t)}\exp\left\{L_{ig}^{(t)}\right\}}{\sum_{g=1}^G\pi_g^{(t)}\exp\left\{L_{ig}^{(t)}\right\}},\\
\tc_{ig}^{(t+1)}=&\frac{\tau_g^{(t)}\exp\left\{L(\bs \xi_{ig1})^{(t)}\right\}}{\tau_g^{(t)}\exp\left\{L(\bs \xi_{ig1})^{(t)}\right\}+\left(1-\tau_g^{(t)}\right)\exp\left\{L(\bs \xi_{ig0})^{(t)}\right\}}.
\end{split}
\end{equation*}

\hspace{-0.5cm}(b) We then update $\pi_g$  and $\tau_g$ as
\begin{equation*}
\begin{split}
& \pi_{g}^{(t+1)}=\frac{1}{n}\displaystyle\sum_{i=1}^n \tz_{ig}^{(t+1)},\quad  \tau_{g}^{(t+1)}=\frac{\displaystyle\sum_{i=1}^n \tz_{ig}^{(t+1)}\tc_{ig}^{(t+1)}}{\displaystyle\sum_{i=1}^n \tz_{ig}^{(t+1)}}.
\end{split}
\end{equation*}
When the MLTCN models are used for detecting extreme patterns, $(1-\tau_g)$ represents the percentage of extreme observations and the proportion of normal observations is at least equal to a pre-determined value $\tau^*_g$ (i.e., $\tau^ *_g=0.5$). In this case, we perform a numerical search of the maximum $\tau_{g}^{(t+1)}$ using the {\tt optimize()} function, over the interval $(\tau^*_g, 1)$, of the function 
\begin{equation*}
\sum_{i=1}^n \tz_{ig}^{(t+1)}\left\{\tc_{ig}^{(t+1)}\log \tau_g+\left(1-\tc_{ig}^{(t+1)}\right)\log (1-\tau_g)\right\}.
\end{equation*}
Herein, we use this approach to update $\tau_g$ and we take $\tau ^*_g=0.5$ for $g=1,\ldots, G$.\\

\hspace{-0.5cm}(c) Estimate the likelihood: We approximate the posterior density\\ $p(\by_{ig}|\bx_i, \tz_{ig}^{(t+1)}=1)$ by its variational lower bound $\underline{p}(\by_{ig}|\bx_i, \tz_{ig}^{(t+1)}=1, \boldsymbol{\xi}_{ig}^{(t)} )$, which is a $\text{MVN}(\bs \mu_{ig}^{(t+1)},\bs \Sigma_{ig}^{(t+1)})$ density, where
{\small\begin{equation*}
\begin{split}
&\E\{\text{Cov}(\bY_{ig})|\tc_{ig}^{(t+1)}=1\}=\left\{\bI-2\sum_{m=1}^M B(\xi_{img1}^{(t)}) \bw_{mg}^{(t)} (\bw_{mg}^{(t)})' \right\}^{-1}\equalscolon\bs \Sigma_{ig1}^{(t+1)},\\
&\E(\bY_{ig}|\tc_{ig}^{(t+1)}=1)=\bs \Sigma_{ig1}^{(t+1)}\left\{\sum_{m=1}^M\left(\tx_{im}-\frac{1}{2}+2B(\xi_{img1}^{(t)})\alpha_{mg}^{(t)}\right)\bw_{mg}^{(t)},\right\}\equalscolon\bs \mu_{ig1}^{(t+1)},\\
&\E\{\text{Cov}(\bY_{ig})|\tc_{ig}^{(t+1)}=0\}=\left\{\frac{1}{\eta_g^{(t)}}\bI-2\sum_{m=1}^M B(\xi_{img0}^{(t)}) \bw_{mg}^{(t)} (\bw_{mg}^{(t)})' \right\}^{-1}\equalscolon\bs \Sigma_{ig0}^{(t+1)},\\
&\E(\bY_{ig}|\tc_{ig}^{(t+1)}=0)=\bs \Sigma_{ig0}^{(t+1)}\left\{\sum_{m=1}^M\left(\tx_{im}-\frac{1}{2}+2B(\xi_{img0}^{(t)})\alpha_{mg}^{(t)}\right)\bw_{mg}^{(t)},\right\}\equalscolon\bs \mu_{ig0}^{(t+1)},
\end{split}
\end{equation*}}
where $B(\xi_{img}^{(t)})=\left\{\frac{1}{2}-\sigma(\xi_{img}^{(t)})\right\}/2\xi_{img}^{(t)}$ and $\sigma(\xi_{img}^{(t)})=\left\{1+\exp(-\xi_{img}^{(t)})\right\}^{-1}$.\\

\hspace{-0.5cm}(d) CM steps 1: Optimize the variational parameter $\xi_{img}^{(t+1)}$. Owing to the EM formulation, each update for $\xi_{img}$ corresponds to a monotone improvement to the posterior approximation. The updates are
{\small\begin{equation*}\begin{split}
(\xi_{img1}^{2})^{(t+1)}&=(\bw_{mg}^{(t)})'\left\{\bs \Sigma_{ig1}^{(t+1)}+\bs \mu_{ig1}^{(t+1)} \bs (\mu_{ig1}^{(t+1)})'\right\}\bw_{mg}^{(t)}+2\alpha_{mg}^{(t)}(\bw_{mg}^{(t)})' \bs \mu_{ig1}^{(t+1)}+(\alpha_{mg}^{(t)})^2,\\
(\xi_{img0}^{2})^{(t+1)}&=(\bw_{mg}^{(t)})'\left\{\bs \Sigma_{ig0}^{(t+1)}+\bs \mu_{ig0}^{(t+1)} \bs (\mu_{ig0}^{(t+1)})'\right\}\bw_{mg}^{(t)}+2\alpha_{mg}^{(t)}(\bw_{mg}^{(t)})' \bs \mu_{ig0}^{(t+1)}+(\alpha_{mg}^{(t)})^2.
\end{split}\end{equation*}}\\

\hspace{-0.5cm}(e) CM step 2: Update $\alpha_{mg}$ and $\bw_{mg}$ based on the posterior distributions corresponding to the observations in the data set:
{\scriptsize\begin{equation*}
\begin{split}
\hat{\bw}_{mg}^{(t+1)}=&
-\left[2\sum_{i=1}^n \tz_{ig}^{(t+1)}\left\{\tc_{ig}^{(t+1)}B(\xi_{img1}^{(t+1)})\E(\bY_{ig1}\bY_{ig1}')^{(t+1)}\right\}\left\{(1-\tc_{ig}^{(t+1)})B(\xi_{img0}^{(t+1)})\E(\bY_{ig0}\bY_{ig0}')^{(t+1)}\right\}\right]^{-1}\\
&\times \left[\sum_{i=1}^n \tz_{ig}^{(t+1)}(\tx_{im}-1/2)\left\{\tc_{ig}^{(t+1)}\hat{\bs \mu}_{ig1}^{(t+1)}+(1-\tc_{ig}^{(t+1)})\hat{\bs \mu}_{ig0}^{(t+1)}\right\}\right],
\end{split}
\end{equation*}}
\hspace{-0.15cm}where $\hat{\bw}_{mg}^{(t+1)}=\{(\bw_{mg}^{(t+1)})', \alpha_{mg}^{(t+1)}\}'$,   $\hat{\bs \mu}_{ig}^{(t+1)}=\{\bs (\mu_{ig}^{(t+1)})', 1\}'$, and for $k=0$ or $1$
\[ \E(\bY_{igk}\bY_{igk}') = \left[ \begin{array}{cc}
\bs\Sigma_{igk}^{(t+1)}+\bs \mu_{igk}^{(t+1)} (\bs \mu_{igk}^{(t+1)})' & \bs \mu_{igk}^{(t+1)}  \\
\bs \mu_{igk}^{'(t+1)}& 1  \end{array} \right].\]

\hspace{-0.5cm}Update $\eta_g$ by optimizing the following log likelihood with respect to $\eta_g$ and subject to $\eta_g>1$,
{\small\begin{equation*}
-\frac{D}{2}\sum_{i=1}^n \left\{\tz_{ig}^{(t+1)}\left(1-\tc_{ig}^{(t+1)}\right)\log \eta_g\right\}-\frac{1}{2}\sum_{i=1}^n \tz_{ig}^{(t+1)}\left(\frac{1-\tc_{ig}^{(t+1)}}{\eta_g}\right)\E(\bY_{ig0}'\bY_{ig0})^{(t+1)},
\end{equation*}}
where $\E(\bY_{ig0}'\bY_{ig0})^{(t+1)}=\Tr\{\E(\bY_{ig0}'\bY_{ig0})^{(t+1)}\}=\Tr\{\E(\bY_{ig0}\bY_{ig0}')^{(t+1)}\}$.\\

\hspace{-0.5cm}(f) Obtain the lower bound of the log likelihood at the expansion point $\boldsymbol \xi_{ig}$ 
{\small\begin{equation*}
\begin{split}
L(\bs \xi_{ig1})^{(t+1)}=& \sum_{m=1}^M \left\{\log\sigma(\xi_{img1}^{(t+1)})-\frac{\xi_{img1}^{(t+1)}}{2}-B(\xi_{img1}^{(t+1)})(\xi_{img1}^{(t+1)})^2\right\}\\
&+\frac{1}{2}\log|\bs\Sigma_{ig1}^{(t+1)}|+\frac{1}{2}(\bs \mu_{ig1}^{(t+1)})'\bs \Sigma_{ig1} ^{-1(t+1)}\bs \mu_{ig1}^{(t+1)},\\
L(\bs \xi_{ig0})^{(t+1)}=&\sum_{m=1}^M \left\{\log\sigma(\xi_{img0}^{(t+1)})-\frac{\xi_{img0}^{(t+1)}}{2}-B(\xi_{img0}^{(t+1)})(\xi_{img0}^{(t+1)})^2\right\}\\
&+\frac{1}{2}\log|\bs\Sigma_{ig0}^{(t+1)}|+\frac{1}{2}(\bs \mu_{ig0}^{(t+1)})'\bs \Sigma_{ig0} ^{-1(t+1)}\bs \mu_{ig0}^{(t+1)},
\end{split}
\end{equation*}} Then, 
\begin{equation*}
l^{(t+1)}\approx \sum_{i=1}^n\log\left[\sum_{g=1}^G\pi_g^{(t+1)} \exp\left\{L_{ig}^{(t+1)}\right\}\right],
\end{equation*} where {\small$L_{ig}^{(t+1)}=\log\left[\tau_g^{(t+1)}\exp\left\{L(\bs \xi_{ig1})^{(t+1)}\right\}+\left(1-\tau_g^{(t+1)}\right)\exp\left\{L(\bs \xi_{ig0})^{(t+1)}\right\}\right].$}


\end{document}